\documentclass[a4paper]{jpconf}
\usepackage{graphicx}
 \usepackage{amsmath}
 \usepackage{amssymb}
  \usepackage{enumerate}
\usepackage{cite}
\usepackage{mathabx}
\usepackage{url}
\newcommand{\bey}{\begin{eqnarray}}
\newcommand{\eey}{\end{eqnarray}}
\usepackage{etoolbox}

\usepackage{lipsum}

\let\OLDthebibliography\thebibliography
\renewcommand\thebibliography[1]{
  \OLDthebibliography{#1}
  \setlength{\parskip}{0pt}
  \setlength{\itemsep}{2pt plus 0.3ex}
}

\begin{document}
\title{Towards a Field-Theory based \\ Relativistic Quantum Information}
\author {Charis Anastopoulos $^{(a)}$,
Bei-Lok Hu $^{(b)} $,
and   Konstantina Savvidou $^{(a)}$}

\address {\small $^{(a)}$ Department of Physics, University of Patras, 26500 Greece}
\address{ $^{(b)}$  Maryland Center for Fundamental Physics and Joint Quantum Institute,
 University of Maryland, College Park, Maryland 20742-4111 U.S.A.}

\ead{anastop@upatras.gr, blhu@umd.edu, ksavvidou@upatras.gr  }

\begin{abstract}
We present our program for the development of quantum informational concepts in relativistic systems in terms of the unequal-time correlation functions of quantum fields. We employ two formalisms that provide the basis for further developments. (i) The Quantum Temporal Probabilities (QTP) Method for quantum field measurements and (ii) the Closed-Time-Path (CTP) formalism for causal time evolutions. We present the main ideas of QTP and  show how it relates to the CTP formalism, allowing one to express concepts of measurement theory in terms of path-integrals. We also present many links of our program to non-equilibrium quantum field theories.  Details can be found in a recent paper by the authors \cite{AHS23}.
\end{abstract}

\section{Introduction}

Quantum information theory (QIT) is a quantum extension of classical information theory.  It has identified new and powerful informational resources for quantum computing,  quantum communication, quantum metrology and more. Despite the broad domain of applicability of  quantum information technologies, our understanding of QIT is far lagging behind the fully developed quantum theory of nature, namely, quantum field theory (QFT). QFT which has proven its validity and worth in the full range of physical sciences from particle-nuclear physics to atomic, optical and condensed matter physics, from quarks and nucleons to black holes and the early universe.  So far, quantum information theory has been largely developed in the context of non-relativistic quantum mechanics, which is a small corner of full fledged QFT.  It is ostensibly inadequate when basic relativistic effects like locality, causality and spacetime covariance, need be accounted for. Recognizing the importance of these relativistic effects and seeking to understand the essential roles they play in quantum information ushered in the emergent field of relativistic quantum information (RQI) \cite{RQI}.

\subsection{QITs not based on QFT are incomplete}

QFT satisfies the principles of quantum theory, but in addition, it is constrained by axioms that govern the effects of the  spacetime structure on the causal behavior of quantum systems. Such axioms are largely missing from  current quantum information theories. The latters' notion of causality, based on the sequence of successive operations on a quantum system, lacks a direct spacetime representation. As a result, current QITs   cannot make crucial relativistic distinctions, for example, between timelike and spacelike correlations, they  do not describe real-time signal propagation, and they ignore relativistic constraints on physical operations. A genuinely relativistic QIT must overcome such limitations \cite{AHS23, AnSav22}.

Furthermore, experiments that study causal information transfer or gravitational interactions in multi-partite quantum systems require a QFT treatment of interactions for consistency.  A non-QFT description is likely to misrepresent either the theoretical modeling of a system or the physical interpretation of the results. This point is crucial for tests of foundational issues of quantum theory invoking quantum information concepts such as entanglement and decoherence. This is especially so for quantum  information  experiments in space \cite{Rideout, DSQL2} and for experiments designed to explore quantum effects from gravity \cite{AnHu15, Bose17, Vedral17, AnHu20}.

The introduction of the key concepts of spacetime covariance and causality in  QIT forces us to address problems that originate from the foundations of QFT. These include the following,

\medskip

\noindent {\em Quantum States.} In set-ups that involve more than two quantum measurements, the standard  state-update rule implies that the
quantum state is genuinely different when recorded from different Lorentz frames  \cite{PeTe}. There is no problem with the theory’s physical predictions that are expressed in terms of (multi-time) probabilities \cite{WH65}. However, the usual notions of quantum information (entropy, entanglement) are defined through the quantum state, and as such, they are ambiguous in relativistic measurement set-ups.

\medskip

\noindent {\em Local Operations.}   It is a challenging problem to formalize the notion of a localized quantum system in QFT. There are powerful theorems  demonstrating that even  unsharp localization in a spatial region leads to faster-than-light signals \cite{Malament,  Heg1}.  Hence, expressing the crucial quantum informational notion of a local operation in terms of spatial localization can lead to conflicts with relativistic causality.

\medskip

\noindent {\em Projective Measurements.}  There are strong arguments that ideal (i.e., projective) measurements in QFT are incompatible with causality \cite{Sorkin, BJK}, essentially because they change the quantum state over a full Cauchy surface. However, existing quantum information notions (including the very notion of a qubit) presuppose maximal extraction of information through ideal measurements.

\medskip

We contend that the development of consistent relativistic QIT requires a measurement theory that (i) respects causality and locality, and (ii) it is expressed in terms of quantum fields.  Furthermore, this measurement theory ought to be practical, i.e., it must provide non-trivial predictions for experiments that are accessible now or in the near future.


\subsection{Past work on QFT measurements}

The earliest discussion of measurements on quantum fields was by Landau and Peierls \cite{LP31}, who  derived a bound to particle localization. Bohr and Rosenfeld criticized their work \cite{BoRo}, and proved that the measurement of definite field properties  requires macroscopic test particles: the charge $Q$ of a test particle must be much larger than the electron charge e.

The first explicit model for QFT measurements was  Glauber's photodetection theory \cite{Glauber1, Glauber2}  that provided a foundation for the then nascent field of quantum optics. This theory involves unnormalized probabilities for photon detection in terms of the electric field operators $\hat{\bf E}(x) $  and the field's quantum state $|\psi\rangle$. The joint probability density $P(x_1, x_2, \ldots, x_n)$ for the detection of a photon at each of the spacetime points $x_1, x_2, \ldots, x_n$ is given by
\bey
P(x_1, x_2, \ldots, x_n)=  \langle \psi| \hat{E}^{(-)}(x_1)\hat{E}^{(-)}(x_2)  \ldots \hat{E}^{(-)}(x_n) \hat{E}^{(+)}(x_n) \ldots \hat{E}^{(+)}(x_2) \hat{E}^{(+)}(x_1)|\psi\rangle, \label{jointdet1}
\eey
where $\hat{E}^{(+)}$ is the positive-frequency component  and $\hat{E}^{(-)}$ the negative-frequency component of the projected field vector field ${\bf n}\cdot \hat{\bf E}(x)$.
The probability density (\ref{jointdet1})  is essential for the definition of high-order correlations of the electromagnetic field, and consequently, for the description of phenomena like the Hanbury-Brown-Twiss effect,   photon bunching and anti-bunching \cite{QuOp}.
Glauber's theory has been highly successful, but  its scope is limited in that it only applies to the quantum electromagnetic field. Furthermore,  it may face causality problems in set-ups that involve the  propagation of photons over long distances, because   the field splitting   into positive and negative frequencies
misrepresents retarded propagation.

Perhaps the simplest models for QFT measurements can be constructed using the notion of an {\em Unruh-DeWitt} (UDW)  detector  \cite{Unruh76, Dewitt}. UDW detectors first appeared in the study of the Unruh effect, where they were employed in order to demonstrate the effects of acceleration on the quantum field vacuum. In a UDW detector model, the quantum field is coupled to a point-like   system that  moves along a pre-determined spacetime trajectory. UDW detectors have found several applications---see, for example, Ref.  \cite{HLL12}---besides their use as models for QFT measurements \cite{GGM22}. They are limited in that the  detector degrees of freedom are not described by  a QFT, a problem that may lead to  non-causal signals in set-ups with multiple detectors.

 Measurement models have also been constructed in the context of  algebraic QFT \cite{HeKr, OkOz, Dop, FeVe}. The idea is to consider a measured    system and a probe / apparatus that are both described by a QFT. The two field systems start separated and   interact within a bounded spacetime region. In Minkowski spacetime, this interaction is described by an S-matrix, and it leads to correlations between observables on the system and records on the probe.
  Then, one defines probabilities for the latter in terms of operators that are well defined on the probe's Hilbert space.   This approach is fully consistent with QFT, it works for curved spacetimes even in absence of asymptotic in-out regions,
   but it has not  yet been developed into a practical tool capable of concrete physical predictions.

\subsection{The Quantum Temporal Probabilities approach}

Here we present an approach toward QFT measurements in terms of the Quantum Temporal Probabilities (QTP) method \cite{QTP1, QTP2, QTP3,  QTP5}. The name of this method is due to its original  motivation to provide a general framework for  temporally extended quantum observables \cite{AnSav06, AnSav08, An08}.

The key idea  in QTP is to distinguish between the time parameter of Schr\"odinger's equation from the time variable associated to  particle detection \cite{Sav99, Sav10}. The latter is then treated as  a macroscopic quasi-classical variable associated to the detector degrees of freedom. A quasi-classical variable is a coarse-grained quantum variable that  satisfies appropriate decoherence conditions, so that its time evolution can be well approximated by  classical equations  \cite{GeHa2, Omn1}.  Hence, the detector admits a dual description: in microscopic scales it is described   by quantum theory, but its macroscopic records are expressed  in terms of classical spacetime coordinates.

In QTP the detector is also described in terms of quantum fields.
 Glauber's detection theory and Unruh-DeWitt detector models emerge from QTP as limiting cases, the former in the limit of very short detector time-scales, the second in the limit of very short detector length-scales \cite{AHS23}.  In comparison to the algebraic QFT approaches to measurements, QTP provides the same results to leading order in perturbation theory, but allows for the definition of observables for the spacetime coordinates, and it is embedded within a nuanced analysis of the quantum-classical transition in the detector.

A key result in QTP is that probabilities for measurements are expressed in terms of specific unequal-time field
correlation functions. Such correlation functions are a staple of QFT.  Powerful  methods  have been developed for their calculation and the analysis of their properties. The specific correlation functions relevant to QTP   appear in the Closed-Time-Path (CTP) (Schwinger-Keldysh or `in-in') formalism \cite{ctp1, ctp2, ctp3, ctp4, Jordan, CH88}. The CTP formalism improves over the S-matrix (in-out formalism), in that it allows for causal equations of motion, and it has found many applications in  nuclear-particle process \cite{CH08, Berges, Berges2}, early universe cosmology \cite{CH87, wein05}, and condensed matter physics \cite{coma1, coma2}.
We demonstrate the link between the two formalisms, and this  allows us to translate between the concepts  of quantum measurement theory and of quantum field theory.

 In QTP, unequal-time correlation functions contain all information about measured probabilities. In particular, the detection probability for $N$ events is a linear functional of a specific $2N$-unequal time correlation function. This has the following implication. For $N = 2$, probabilities of measurement outcomes are related to bipartite entanglement. Hence, all information about bipartite entanglement  is contained in the field four-point functions.
    
    An analysis at the level of the correlation functions brings us closer to the main ideas of non-equilibrium QFT. Indeed, we can establish  a natural relation between QTP and non-equilibrium formalisms that are based on CTP.  QTP probabilities function as a registrar of information for the quantum field, they keep track  of how much information resides in which level of correlation functions, and how this information flows from one level to the other during dynamical evolution.

We believe that the scheme outlined here has good potential to systemize quantum information in QFT,   and to identify the parts of this information that is relevant to the field's statistical, stochastic and thermodynamic behavior.
  Hence, this formalism could provide a concrete method for defining  quantum information in QFT via the correlation hierarchy, as has been proposed in Ref. \cite{Erice95}. Such a definition would be very different from definitions of information in standard QIT that is based on the properties of the single-time quantum state.

\section{Probabilities for QFT measurements }
We first explain the need for a QFT measurement theory, and then present the  QTP approach to such measurements.  The key property of the QTP probability formula is that the probability density for $n$ measurement events is a linear functional of a specific $2n$ unequal-time correlation function.

\subsection{Why we need a QFT measurement theory}

Most current applications of QFT involve  the S-matrix formalism. For example,  S-matrix amplitudes determine scattering cross-sections;  S-matrix poles determine the spectrum of composite particles and decay rates. S-matrix theory is defined for set ups with a single state preparation and to a single detection event in the asymptotic future. This gives the impression that there is no need for an elaborate measurement theory.

This impression is wrong, because there are at least two cases, where the S-matrix formulation of QFT does not suffice.   First, in  quantum optics, we need joint detection probabilities  in order to describe phenomena like photon bunching and anti-bunching \cite{QuOp}. In non-relativistic physics, joint probabilities of this type involve the use of the state-update rule, i.e., quantum state reduction. However, a universal rule for reduction is missing in QFT. In practice,
 joint detection probabilities relevant to experiments are constructed through heuristic arguments, for example, as, for example, in Glauber's  photodetection models.
  Planned experiments in deep space \cite{Rideout,  DSQL2} that involve  measurement of quantum optical correlations at long distances  arguably require a first-principles construction of joint probabilities.

Second, S-matrix is insufficient whenever we are interested in expectation values of physical quantities at finite moments of time, rather than scattering amplitudes \cite{ctp4, Jordan}. Examples include the description of     quantum transport in  many-body systems \cite{CH88, Berges, CH08}, and the backreaction of quantum fields on the spacetime metric in  cosmological and black hole spacetimes \cite{CH87, wein05}.  Powerful functional techniques, like the Schwinger-Keldysh method, have been developed to deal with such problems.

\subsection{The Quantum Temporal Probabilities Approach: main ideas}

The key features of the QTP approach to measurements on quantum fields are the following.
\begin{enumerate}
\item The apparatus is  fundamentally described in terms of QFT. In particular, the interaction between the measured system and the apparatus is described by a Hamiltonian that is a local functional of quantum fields.

\item The measurement apparatus is also assumed to exhibit classical behavior at the macroscopic level.   According to the decoherent histories approach to emergent classicality \cite{Omn1, GeHa2}, the apparatus pointer is a highly coarse-grained observable, so that histories for measurement outcomes satisfy appropriate decoherence conditions.

\item All measurements events are localized in space and in time. For example, a particle detector has a fixed location in a laboratory, and it records an event at a specific moment of time. Both the locus and the time of detection are random variables.  Hence, physical predictions are expressed in terms of probability densities
        \bey
        P(x_1, q_1; x_2, q_2, \ldots, x_n, q_n), \label{probdengen}
        \eey
for multiple detection events. In Eq. (\ref{probdengen}), $x_i$ stand for spacetime points, $q_i$ stand for any other recorded observable and $P$ is a probability density with respect to both $x_i$ and $q_i$.

\end{enumerate}

 \subsection{Detection probability for a single detector}
Consider a QFT on Minkowski spacetime $M$: it is described by  Heisenberg-picture fields $\hat{\phi}_r(x)$ that are defined on a Hilbert space ${\cal F}$. The Hilbert space carries
 a unitary representation of the Poincar\'e group, and  The index $r$  runs over spacetime and internal indices.

Let ${\cal K}$ be
the Hilbert space associated to an apparatus. We assume that the apparatus follows a world tube ${\cal W}$ in Minkowski spacetime, and that its size
 is   much larger than the scale of microscopic dynamics.
We introduce a field-apparatus coupling with support in a small spacetime region around a point $x$. The finite spacetime extent of the interaction  mimics the effect of a detection record localized at $x$. Working in the interaction picture, we express the coupling term as
\bey
\hat{V}_x = \int F_x(y) \hat{C}_a(y) \otimes \hat{J}^a(y), \label{VX}
\eey
where $\hat{C}_a(x)$ is a composite operator on ${\cal F}$ that is local with respect to the field $\hat{\phi}_r(x)$ and $a$ runs over spacetime and internal indices. The current operators $\hat{J}^a(x)$ are defined on ${\cal K}$.
 The switching functions $F_x(y)$ are dimensionless. They vanish outside the interaction region and they depend on the motion of the apparatus. The spacetime volume  associated to a switching function is $\upsilon = \int dY F^2_x(y)$.

The switching function renders the interaction term (\ref{VX})  Poincar\'e non-covariant. The use of switching functions originates from von Neumann's modeling of quantum measurements \cite{vN}, where it serves to localize the system-apparatus interaction in time. Certainly, in models with a switching functions, time is not a random variable. A switching function is not needed in QTP, but we employ it in ths work, because it is computationally easier, and it leads to the same expressions for probabilities to leading order in perturbation theory.

The S-matrix associated to Eq. (\ref{VX}) is $\hat{S}_x = {\cal T} \exp[ - i \int d^4y F_x(y) \hat{C}_a(y) \otimes \hat{J}^a(y)]$, where ${\cal T}$ stands for time ordering. To leading order in the interaction,
\bey
\hat{S}_x = \hat{I} - i \hat{V}_x.
\eey
Let the initial state of the system  be $|\psi\rangle \in {\cal F}$ and the initial state of the apparatus  be $|\Omega\rangle$. A particle record appears if the detector transitions from $|\Omega\rangle$ to its complementary subspace ${\cal K}'$. Once the transition occurred, we  measure a property of the particle through a pointer observable $q$. The latter is described by a family of positive operators $\hat{\Pi}(q)$, such that $\sum_{q} \hat{\Pi}(q) = \hat{I} - |\Omega \rangle \langle \Omega|$. The pointer observable is coarse-grained, and we take it to stationary with respect to the  self-dynamics of the detector, so that the record is preserved after the end of the measurement.

Then, we compute the probability $\mbox{Prob}(x, q)$ that the detector is excited and records a value $q$,
\bey
\mbox{Prob}(x, q) = \langle \psi, \Omega| \hat{S}^{\dagger}_x[\hat{I} \otimes \hat{\Pi}(q)]\hat{S}_x|\psi, \Omega\rangle
\eey
To leading order in perturbation theory
\bey
\mbox{Prob}(x, q) = \int d^4y_1 d^4y_2  F_x(y_1) F_x(y_2) G_{ab}(y_1, y_2) \langle \Omega|\hat{J}^a(y_1) \hat{\Pi}(q) \hat{J}^b(y_2)|\Omega\rangle, \label{probX}
\eey
where
\bey
G_{ab}(x, x')  = \langle \psi|\hat{C}_a(x) \hat{C}_b(x')|\psi\rangle,
\eey
is a correlation function for the composite operator.

The probability $\mbox{Prob}(x, q)$ of Eq. (\ref{probX}) is not a density with respect to $x$, because $x$ appears as a parameter of the switching function.   We define an unnormalized probability density $W(x, q)$ with respect to $x$ by dividing $\mbox{Prob}(x, q)$ with the effective spacetime volume $\upsilon$,
\bey
W(x, q) = \upsilon^{-1}\mbox{Prob}(x, q). \label{Prob0b}
\eey

The  definition (\ref{Prob0b})   is well justified classically, but not in quantum theory. It combines probabilities defined with respect to different experimental set-ups, i.e., different switching functions, a practice that is frowned upon by quantum complementarity.
Nonetheless,   Eq. (\ref{Prob0b}) can  be derived as a genuine probability density through the QTP method \cite{QTP3, AHS23}, to leading order in the field-apparatus coupling.

   In the  proper QTP derivation, the interaction is present at all times, as the Hamiltonian is time-translation invariant.
The   functions $F_x(y)$ are not   switching functions of the interaction,
 but they describe the {\em sampling} of the spacetime point. Hence, the spacetime volume $\hat{\upsilon}$ is a measure of {\em coarse-graining}, i.e., of the inaccuracy in the determination of the spacetime point. This point is crucial for a rigorous derivation, because probabilities can only be defined   for histories that satisfy a decoherence condition, for which coarse-graining is a prerequisite. In principle,  the minimum  coarse-graining scale compatible with well-defined probabilities is determined from first-principles, so $\sigma$ is not arbotrary---see \cite{QTP1} for explicit calculations in simple models.
  This means that not all sampling functions $F_x(y)$ are acceptable:  their support  cannot be made arbitrarily small. Such constraints cannot be seen in the derivation presented here.

To further proceed, we assume that the detector carries a representation of the spacetime translation group with generators $\hat{p}^{\mu}$.  We note that the state $|\Omega\rangle$ is not the Poincar\'e invariant vacuum; it defines a preferred reference system at which the expectation of the apparatus's total momentum vanishes.

We choose a reference point $x_0 $ in  the detector's world-tube, and we write
\bey
\hat{J}^{a}(y) = e^{-i \hat{p} \cdot (y - x_0)} \hat{J}^a(x_0) e^{i \hat{p} \cdot (y - x_0)}.
\eey
It is convenient to take
$|\Omega\rangle$ to be   {\em approximately translation invariant}, i.e., to require that
\bey
\int d^4 x F_{x}(x') \hat{J}^a(x') |\Omega\rangle \simeq \int d^4 x F_{x}(x') e^{-i \hat{p} \cdot (x' - x_0)} \hat{J}^a(x_0)|\Omega\rangle.
\eey
 The idea is that the apparatus is prepared in an initial state that is homogeneous at the length scales that correspond to position sampling and approximately static at the time scales that correspond to time sampling.
With this assumption,   $\langle \Omega|\hat{J}^a(y_1) \hat{\Pi}(q) \hat{J}^b(y_2)|\Omega\rangle = R^{ab}(y_2 - y_1, q)$,
 where
\bey
R^{ab}(x, q) := \langle a, q|e^{i \hat{p} \cdot (x - x_0)}|b, q\rangle \label{detkern}
\eey
is the {\em detector kernel}, expressed in terms of the vectors
 $|a, q\rangle = \sqrt{\hat{\Pi}}(q) \hat{J}^a(x_0)|\Omega\rangle$. Note that the Fourier transform of the detector kernel,
$\tilde{R}^{ab}(\xi, q) = \int d^4x e^{-i \xi\cdot x} R^{ab}(x, q)$
is given by
\bey
\tilde{R}^{ab}(\xi, q) = (2\pi)^4 e^{i \xi\cdot x_0} \langle a, q|\hat{E}_{\xi}|b, q\rangle,
\eey
where $\hat{E}_{\xi} = \delta^4(\hat{p} - \xi)$ is the projector onto the subspace with four-momentum $\xi^{\mu}$.

The simplest  switching functions $F_x$ are Gaussians, of the form $F_x(y) = f(x-y)$, where
\bey
f(x) = \exp[ - \frac{1}{2 \delta_t^2} (x^0)^2 -\frac{1}{2 \delta_x^2} {\bf x}^2],
\eey
where $\delta_t$ is the  temporal accuracy and $\delta_x$ is the special accuracy of the detector. These quantities are macroscopic, because they correspond to the sampling of the detection event.

 Gaussian switching functions satisfy the identity
\bey
f(x) f(x') = f^2\left(\frac{x+x'}{2}\right) \sqrt{f}(x - x'). \label{gauidty}
\eey
The spacetime volume $\upsilon$ of the interaction region  is $\upsilon =\pi^2 \delta_t \delta_x^3$. We note that the function   $\sigma(x): = \frac{1}{\upsilon} f^2(x)$ is a normalized probability density on $M$. Then, we write
\bey
W(x, q) = \int d^4x'  \sigma(x - x') P(x', q), \label{probX20}
\eey
where
\bey
P(x, q) = \int d^4y  \sqrt{f}(y) R^{ab}(y, q) G_{ab}(x - \frac{1}{2}y, x +\frac{1}{2}y), \label{probX2}
 \eey

The probability distribution $W(x, q) $ is the convolution of $P(x, q)$ with the probability density $\sigma(x)$ that accounts for the accuracy of our measurements. If $P(x, q)$ is non-negative and the scale of variation with respect to $x$ is much larger than both $\delta_t$ and $\delta_x$, we can treat $P(x, q)$  as a fine-grained version of $W(x, q)$ and use this as the probability density for detection.

The kernel $R^{ab}(x, q)$ is typically characterized by a correlation length-scale $\ell$ and a correlation time-scale $\tau$, such that $ R^{ab}(x, q)  \simeq 0$ if $|t(\xi)| \gg \tau $ or  $|{\bf x}(\xi)|\gg \ell$.
Both scales $\ell$ and $\tau$ are microscopic and characterize the constituents of the apparatus and their dynamics. If $\ell \ll \delta_x$ and $\tau \ll \delta_t$, then $R^{ab}(x, q) \sqrt{f}(x) \simeq R^{ab}(x, q)$ and we obtain an expression for the probability density  $P(x, q)$ that is sampling-independent
\bey
P(x, q) = \int d^4 y     R^{ab}(y, q) G_{ab}(x - \frac{1}{2}y, x +\frac{1}{2}y). \label{prob1aa}
\eey
The probability densities (\ref{prob1aa}) are not normalized to unity. In general, the total probability of detection $P_{det} = \sum_{q} \int_{\cal W} d^4x  P(q, x) $ is  a small number in any perturbative calculations. There is always a non-zero probability $P(\emptyset) = 1 - P_{det}$ of no detection. We normalize probabilities  by conditioning the probability densities $P(q, x)$ with respect to the existence of a detection record. Hence, we use the probability densities  $P(x, q)/P_{det}$.

\subsection{Multiple detectors}
It is straightforward to obtain the probability formula for the field interacting with  $n$ detectors.
To leading order in perturbation theory,  the probability density for $n$ measurement events is
\bey
W_n(x_1, q_1; x_2, q_2; \ldots; x_n, q_n) = \int d^4x'_1 \ldots d^4 x'_n \sigma^{(1)}(x_1 - x'_1) \ldots \sigma^{(n)}(x_n - x'_n)\nonumber \\ P(x'_1, q_1;  x'_2, q_2; \ldots; x'_n, q_n)
\eey
where
\bey
P_n(x_1, q_1; x_2, q_2; \ldots; x_n, q_n) = \int d^4 y_1 \ldots d^4 y_n   \sqrt{f^{(1)}}(y_1) \ldots \sqrt{f^{(n)}}(y_n)
R_{(1)}^{a_1b_1}(y_1, q_1) \ldots
\nonumber \\
\times \ldots R_{(n)}^{a_nb_n}(y_n, q_n)  G_{a_1 \ldots a_n, b_1\ldots b_n}(x_1 - \frac{1}{2}y_1, \ldots, x_n - \frac{1}{2}y_n; x_1 + \frac{1}{2}y_1, \ldots, x_n - \frac{1}{2}y_n). \hspace{0.2cm} \label{probdenN}
\eey
Here $R^{(i)}(x, q)$ is the measurement kernel for the $i$-th detector. The field correlation function $G_{a_1 \ldots a_n, b_1\ldots b_n}(x_1, \ldots, x_n; x_1', \ldots, x_n')$ is given by
\bey
G_{a_1 \ldots a_n, b_1\ldots b_n}(x_1, \ldots, x_n; x_1', \ldots, x_n') = \langle \psi| {\cal T}^*[\hat{C}_{b_1}^{(1)}(x'_1) \ldots \hat{C}_{b_n}^{(n)}(x'_n) ]
\nonumber \\
\times {\cal T} [\hat{C}_{a_n}^{(n)}(x_n) \ldots \hat{C}_{a_1}^{(1)}(x_1)]|\psi\rangle \label{correl}
\eey
where ${\cal T}^*$ stands for reverse time ordering.

Again,   in the appropriate regime, the probability becomes independent of the switching functions, and equal to
\bey
P_n(x_1,q_1; x_2, q_2; \ldots; x_n, q_n) = \int d^4 y_1 \ldots d^4 y_n     R_{(1)}^{a_1b_1}(y_1, q_1) \ldots R_{(n)}^{a_n1b_n}(y_n, q_n)
\nonumber \\
\times G_{a_1 \ldots a_n, b_1\ldots b_n}(x_1 - \frac{1}{2}y_1, \ldots, x_n - \frac{1}{2}y_n; x_1 + \frac{1}{2}y_1, \ldots, x_n + \frac{1}{2}y_n) \label{probden4}
\eey

It is convenient to express the probability densities (\ref{probden4}) using an abstract notation. We use small Greek indices $\alpha, \beta, \gamma \ldots$ for the pairs  $(x, a)$ where $x$ is a spacetime point and $a$ the internal index for the composite operators $\hat{C}_a$. All indices  in a time-ordered product are upper, and all indices in an anti-time-ordered product are lower. Hence, we write the correlation functions (\ref{correl}) as
\bey
G^{\alpha_1 \alpha_2 \ldots \alpha_n}_{\beta_1 \beta_2 \ldots \beta_n} \nonumber
\eey
Let  $x \in M$ be a spacetime point and $q\in \Gamma$ any other recorded observable. We define the set of elementary
events  $Z:= M \times \Gamma \cup \{\emptyset\}$ by $z$, where $\emptyset$ is the event of no detection. Then, we express the kernel
\bey
\sigma[x - \frac{1}{2}(y+y')] \sqrt{f}(y-y') R^{ab}(y - y', q) \nonumber
\eey
as $R_\alpha^\beta(z)$ where $z \in Z$,  $\alpha$ stands for $(y, a)$, $\beta$ for $(y', b)$. We will use the same symbol for the approximate expression $\delta[x - \frac{1}{2}(y+y')]R^{ab}(y - y', q)$. We employ the Einstein summation convention over Greek indices, in order to denote sum over the discrete index $a$ and spacetime integral.

Then, the probability formula (\ref{probden4}) reads
\bey
P_n(z_1, z_2, \ldots, z_n)  = G^{\alpha_1 \alpha_2 \ldots \alpha_n}_{\beta_1 \beta_2 \ldots \beta_n} \;\; {}^{(1)} R_{\alpha_1}^{\beta_1}(z_1) \; {}^{(2)} R_{\alpha_2}^{\beta_2}(z_2) \ldots \; {}^{(n)} R_{\alpha_n}^{\beta_n}(z_n). \label{probdenr}
\eey

\section{Relation of the QTP approach  to the Closed-Time-Path formalism}

The probability density (\ref{probdenN}) for $n$ measurement events is a linear functional of the $2n$-point unequal-time correlation function (\ref{correl}). This correlation function has
 $n$ time-ordered arguments and $n$ anti-time-ordered arguments.  It does not appear in the usual S-matrix description of QFT; the correlation functions in the S-matrix  description involve only time-ordered arguments. Rather,  the correlation function (\ref{correl}) appears in the Schwinger-Keldysh or `in-in' or Closed-Time-Path (CTP) formalism of QFT.  
 
 Since by now the Schwinger-Keldysh formalism is quite well known and popularly used in many fields of physics we shall not belabour it but refer the reader to some source materials \cite{DCH,cddn,CH99} where this method is used for the exploration of themes relevant to our present discussions, foremost, quantum correlations. How the CTP formalism overcomes the deficiencies of the S-matrix formulation is discussed in many original papers on CTP, e.g., \cite{ctp4,Jordan} --   a short summary can be found in our recent paper \cite{AHS23}.

In the CTP formalism, we couple the field to two different external sources $J^a(x)$ and $\bar{J}^a(x)$, and we define the CTP generating functional
\bey
Z_{CTP}[J, \bar{J}] = \langle \psi_0|\hat{U}^{\dagger}[\bar{J}]\hat{U}[J]|\psi_0\rangle,
\eey
By definition, $Z[J, J] = 1$ and $Z^*[J, \bar{J}] = Z[\bar{J}, J]$. The state $|\psi_0\rangle $ is defined in the distant past, i.e., prior to any time at which
 $J(x)$ has support---it is an {\em in} state.

The CTP generating functional describes correlation functions with $n$ time-ordered and $m$ anti-time-ordered entries,
\bey
G^{n,m}_{a_1 \ldots a_m, b_1\ldots b_n}(x_1, \ldots, x_n; x_1', \ldots, x_n') = \langle \psi_0| {\cal T}^*[\hat{C}_{b_1}^{(1)}(x'_1) \ldots \hat{C}_{b_m}^{(n)}(x'_m) ]
\nonumber \\
\times {\cal T} [\hat{C}_{a_n}^{(i)}(x_n) \ldots \hat{C}_{a_1}^{(1)}(x_1)]|\psi_0\rangle \label{correl8}
\eey
which can standardly be expressed as functional derivatives of $Z_{CTP}[J, \bar{J}]$.
 For a vacuum initial state, the generating functional has a path integral expression
\begin{equation}
\label{R8}  Z_{CTP}[J, \bar{J}]
=\int~D\phi~D\bar{\phi} \; e^{i\{S[\phi]-S[\bar{\phi}]+\int d^4x~[J^a(x)C_a (x)
-\bar{J}(x)\bar{C}_a (x)]\} },
\end{equation}
where $\bar{C}$ is defined a functional of $\bar{\phi}$.

The relation between the QTP description of measurements and the CTP formalism is more transparent, if we use the index notation of Sec. 2.4. For consistency, the sources $J$ have a lower Greek index, and the sources $\bar{J}$ have an upper Greek index. Then, we write the CTP generating functional as
\bey
Z_{CTP}[J, \bar{J}] = \sum_{n, m = 0}^{\infty} \frac{ i^{m-n}}{n!m!}  G^{\alpha_1 \ldots \alpha_n}_{\beta_1\ldots \beta_m} J_{\alpha_1} \ldots J_{\alpha_n} \bar{J}^{\beta_1} \ldots \bar{J}^{\beta_m},
\eey
where $G^{\alpha_1 \ldots \alpha_n}_{\beta_1\ldots \beta_m}$ represents the correlation functions (\ref{correl8}). These
 satisfy
\bey
G^{\alpha_1 \ldots \alpha_n}_{\beta_1\ldots \beta_m}  = i^{n-m} \left(\frac{\partial^{n+m}Z_{CTP}[J, \bar{J}]}{\partial J_{\alpha_1} \ldots \partial J_{\alpha_n} \partial \bar{J}^{\beta_1} \ldots \partial \bar{J}^{\beta_n}}\right)_{J = \bar{J}+0}.
\eey
The probability densities (\ref{probdenr}) involve  {\em balanced} correlation functions, i.e., correlation functions  with an equal number of upper and lower indices. We can construct a generating functional that contains only such functions. The key observation is that such correlations contribute to the sum only through products of the form $J_\alpha \bar{J}^{\beta}$. Hence, the natural source for a diagonal generating functional $Z^d_{CTP}$ that only involves balanced correlation functions is a `tensor' $L_{\alpha}^{\beta}$. We define
\bey
Z^d_{CTP}[L] = \sum_{n=0}^{\infty} \frac{1}{n!}G^{\alpha_1 \ldots \alpha_n}_{\beta_1\ldots \beta_m} L_{\alpha_1}^{\beta_1} \ldots L_{\alpha_N}^{\beta_N}.
\eey

Suppose now that we consider only measurements of a single type, i.e., all detector kernels $R_{\alpha}^{\beta}(z)$ are identical. Then, we can define a moment-generating functional for all probability densities
 (\ref{probdenr}), in terms of sources $j(z)$,
\bey
Z_{QTP}[j] = \sum_{n=0}^{\infty} \sum_{z_1, z_2, \ldots, z_n} \frac{1}{n!}  P_n(z_1, z_2, \ldots, z_n) j(z_1) \ldots j(z_n). \label{zqtp}
\eey
It is straightforward to show that
\bey
Z_{QTP}[j]  = Z^d_{CTP}[R\cdot j], \label{fundamental}
\eey
where $(R\cdot j)_A^B = \sum_{z} R_A^B(z) J(z)$.

Eq. (\ref{fundamental}) is a fundamental relation for quantum measurements in QFT, as it relates the moment generating functional for a hierarchy of measured probability densities to the generating functional of unequal-time correlation functions

It is straightforward to write a path integral expression for $Z^d_{CTP}[L]$ for a vacuum initial state
\bey
Z^d_{CTP}[L] = \int~D\phi~D\bar{\phi} \; e^{iS[\phi]- i S[\bar{\phi}]+\int d^4x d^4x' C_a(x) \bar{C}_b(x') L^{ab}(x, x') }.
\eey
To obtain a simple path integral expression for a broader class of states, we recall that many field initial states can be obtained from the action of an external source $\zeta(x)$ on the vacuum, i.e., they are of the form $|\psi_0\rangle = \hat{U}[\zeta]|0\rangle$, where now we write $\hat{U}[\zeta] = {\cal T} \exp\left[i \int d^4 X \zeta^k(X)\hat{A}_k(X)\right]$ in terms of composite operators $\hat{A}_k(x)$ that differ, in general from $\hat{C}_a(x)$---see \cite{AHS23}  for examples.

Hence, for a quantum state that is obtained from an external source $\zeta_k$, we  write the path integral expression
\bey
Z^d_{CTP}[f, L] = \int~D\phi~D\bar{\phi} \; e^{iS[\phi]- i S[\bar{\phi}] + i  \int d^4x~\zeta^k(x)[ A_k (x)
-\bar{A}_k (x)] + \int d^4x d^4x' C_a(x) \bar{C}_b(x') L^{ab}(x, x') },
\eey
where we must assume that the spacetime support of the kernel $L^{ab}$ is later than the support of $\zeta$ (state preparation is prior to measurement).
By Eq. (\ref{fundamental})
\bey
Z_{QTP}[f, j] = \int~D\phi~D\bar{\phi} \; e^{iS[\phi]- i S[\bar{\phi}] + i  \int d^4x~\zeta^k(x)[ A_k (x)
-\bar{A}_k (x)]}
\nonumber \\
\times e^{ \sum_z \int d^4x d^4x' C_a(x) \bar{C}_b(x') R^{ab}(x, x'; z) j(z)} \label{ZQTP}
\eey
The probability densities for $n$ measurement events are obtained from functional variation of $Z_{QTP}[\zeta, j] $ with respect to $j$ at $ j = 0$
For example, the single-event probability density $\hat{P}_1(x, z)$ of Eq. (\ref{prob1aa}) is given by the path integral
\bey
P_1(x, q) =   \int~D\phi~D\bar{\phi} \left(\int d^4y C_a(x+\frac{1}{2}y) \bar{C}_b(x- \frac{1}{2}y) K^{ab}(y, q) \right)
\nonumber \\
     \times         e^{iS[\phi]- i S[\bar{\phi}] + i  \int d^4x~\zeta^k(x)[ A_k (x)
-\bar{A}_k (x)]}. \nonumber
\eey
Expressions such as the above provide an explicit link between concepts of quantum measurement theory like POVMs and  the practical and highly successful  functional language of QFT. We believe that this link is essential for a local and covariant definition of quantum informational notions in QFT.

\section{Links to non-equilibrium QFT}

In Sec. 2, we saw that the QTP probabilities are linear functionals of balanced correlation functions. The measurements do not probe unbalanced correlation functions. Since the latter include $\langle \hat{C}_a(x)\rangle$, QTP probabilities cannot access mean field information. This limitation is not fundamental. Remember that the operator $\hat{C}(x)$ appears in the interaction term with the apparatus. This restriction means that we cannot use couplings of the form $\int d^4x \hat{C}_a(x) \otimes \hat{J}^a(X)$, in order to directly measure the operator $\hat{C}_a(x)$. At least such measurements are not possible with weak field-apparatus coupling where perturbation theory is applicable.

Suppose, for example that $\hat{C}$ coincides with the field operator $\hat{\phi}$---we drop the index $a$ for simplicity.  Then, single-detector probabilities record only local information about particles. Let the field be  in a state characterized by  a macroscopically large number of particles; then, it can be viewed as a thermodynamic system. Then the single-detector probability essentially coincides with a particle-number density function. If we also measure the recorded particle's momentum $k$, the QTP probability density $P(x, k)$ is an operationally defined version of Boltzmann's distribution function.
By Eq. (\ref{prob1aa}), $P(x, k)$ is a linear functional of the correlation function $G(x, x') = \langle \hat{\phi}(x) \hat{\phi}(x')\rangle$, which is usually taken to satisfy the Baym-Kadanoff equations.

From the above analysis, it follows that  Boltzmann's thermodynamic entropy, defined on a Cauchy surface $\Sigma$,
\bey
S_{B}(\Sigma) = -  \int_{\Sigma} d^3 x d^3k P(x,k) \ln P(x,k) \label{boltzent}
\eey
 is a Shannon-type entropy for  single-detection measurements. This means that quantum informational quantities, defined through measurements, have a direct application to non-equilibrium QFT. Furthermore, $n$-detector QTP probabilities probe higher-order correlation functions of the quantum field, thus allowing an analysis that is not accessible by traditional methods.  We shall highlight some  structural similarities and connections between the QTP analysis and methods of non-equilibrium QFT below.

\subsection{Stochastic correlation dynamics from two-particle irreducible effective action}
As shown in Sec. 3 the generating functional of QTP correlation functions is defined in terms of non-local source terms $L^{ab}(x, x')$. It is structurally similar to the two-particle irreducible  effective action (2PIEA) \cite{CH88,RH97} that has found many applications in non-equilibrium QFT---see, for example, \cite{CH08, Berges, Berges2}.   For  present purposes, it suffices to show the structural framework of the 2PIEA formalism, following the presentation in  \cite{cddn, CH99} to define evolution equations with noise from higher-order correlation functions.

For ease of notation, we use a version of DeWitt's condensed notation, where capital indices $A$ correspond to both the spacetime dependence and the branch of the CTP field (forward or backward in time, $\phi$ or $\bar{\phi}$). Hence, we will be writing $\phi_A$, $C_A$, and so on. The action in the CTP generating functional will be $S[\phi_A] = S[\phi_a] - S[\bar{\phi}_a]$. In the two-particle irreducible representation,   the (two-point) correlation function stands is an independent variable, not a functional of the mean field. Thus there is a  separate source $K^{AB}$ driving $C_A C_B$ over the usual $J^A C_A$ term in the one-particle irreducible representation---see the similarity to Eq. (\ref{ZQTP}).

From the generating functional
\begin{equation}
Z\left[ K^{AB}\right] =e^{iW\left[ K^{AB}\right] }=\int D\phi_A
\;e^{i\left( S+\frac 12K^{AB}C_AC_B\right) }  \label{genfun}
\end{equation}
we have
\begin{equation}
G^{AB}=\left\langle \hat{C}_A\hat{C}_B\right\rangle =2\left. \frac{\delta W}{%
\delta K^{AB}}\right| _{K=0}  \label{greenfun}
\end{equation}
and
\begin{equation}
\left. \frac{\delta^2W}{\delta K^{AB}\delta K^{CD}}\right| _{K=0}=\frac{i}{4}%
\left\{ \left\langle \hat{C}_A \hat{C}_B \hat{C}_C \hat{C}_D \right\rangle -\left\langle
 \hat{C}_A \hat{C}_B \right\rangle \left\langle\hat{C}_C \hat{C}_D\right\rangle \right\}.
\label{fluc}
\end{equation}
Suppose that we want to express the effective dynamics of $G_{AB}$ in a closed form, but to go beyond the Baym-Kadanoff equations, by taking into account noise from higher-order correlations.
For a non-equilibrium system, we seek a  formulation in terms of  a new object  ${\bf G}_{AB}$. This is  a stochastic correlation function whose expectation value over the noise average  gives the usual two point functions. The fluctuations of ${\bf G}_{AB}$ reproduce the quantum fluctuations in the binary products of field operators. The simplest assumption is to take  ${\bf G}_{AB}$ as a Gaussian process, defined
by

\begin{equation}
\left\langle {\bf G}_{AB}\right\rangle =\left\langle \hat{C}_A\hat{C}_B\right\rangle ;\qquad \left\langle {\bf G}_{AB}{\bf G}_{CD}\right\rangle
=\left\langle \hat{C}_A \hat{C}_B \hat{C}_C \hat{C}_D\right\rangle  \label{stocg}
\end{equation}


The Legendre transform of $W$ is the two-particle irreducible effective action,
\begin{equation}
\Gamma_{2PI} \left[ G_{AB}\right] =W\left[ K^{AB}\right] -\frac 12K^{AB}%
{G}_{AB};\qquad K^{AB}=-2\frac{\delta \Gamma }{\delta {G}_{AB}}
\label{tpiea}
\end{equation}
The Schwinger-Dyson equation for the propagators is simply, $\frac{\delta \Gamma_{2PI} }{\delta G_{AB}}=0$. When including the stochastic source  ${\bf G}_{AB}$, it becomes
\begin{equation}
\frac{\delta \Gamma_{2PI} }{\delta {\bf G}_{AB}}=-\frac{1}2\kappa^{AB}
\label{lan2pi}
\end{equation}
where $\kappa _{ab}$ is a stochastic nonlocal Gaussian source defined by
\begin{equation}
\left\langle \kappa^{AB}\right\rangle =0;\qquad \left\langle \kappa^{AB}\kappa^{CD}\right\rangle =4i\left[ \frac{\delta ^2\Gamma_{2PI} }{\delta
G_{AB}\delta G_{CD}}\right] ^{\dagger }  \label{noisecor}
\end{equation}

The noiseless Eq. (\ref{lan2pi}) ($\kappa = 0$) provides the
basis for the derivation of transport equations in the near equilibrium
limit. Indeed, for a $\lambda \phi ^4$ theory, we obtain the Boltzmann equation for a distribution function $f$ defined from the Wigner transform of $G^{ab}$.
The full stochastic equation (\ref{lan2pi}) leads, in the same limit, to a Boltzmann - Langevin equation \cite{CH99}.

\subsection{Correlation Histories}
The two-particle irreducible formalism can be extended to an n-particle irreducible formalism, for any $n$. There is an effective action $\Gamma_{nPI}$  for each $n$, from which all effective actions for $n' < n$ can be derived. Taking $n\rightarrow \infty$, we obtain a master effective action. The functional variation of the master effective action  yields the hierarchy of Schwinger-Dyson equations \cite{cddn}.

To obtain effective closed dynamics for the correlations at order $n$, we must truncate the Schwinger-Dyson hierarchy upon this order. Truncation renders the master effective action complex. Its imaginary part arises from correlation functions of order higher than $n$, the fluctuations of which Calzetta and Hu define as  {\em correlation noises} \cite{CH99} at order $n$. For example, the noise $\kappa^{AB}$ in Eq. (\ref{lan2pi}) is the correlation noise of order two.

Calzetta and Hu defined the notion of {\em correlation histories} \cite{DCH}, in analogy to the decoherent histories program. A fine-grained correlation  history corresponds to the full Schwinger-Dyson hierarchy of correlation functions. When we truncate the hierarchy at finite order $n$, we treat only correlation functions of order $n$ as independent. Higher -order correlations are ignored or slaved to  the lowest-order ones.
A truncated hierarchy defines a {\em coarse-grained} correlation history. For example, mean field theory studies {\em coarse-grained} correlation histories at order $n = 1$; the Baym-Kadanoff equation, or Boltzmann equation and their stochastic generalizations  refer to {\em coarse-grained} correlation histories of order $n = 2$.

The key point is that the truncation of the master effective action always leads to dissipation and noise for the coarse-grained histories. Any truncated theory is an effective field theory in the correlation hierarchy formulation. This effective field theory does not carry the full information, this loss of information being expressed as   correlation noise.
The higher-order correlations are analogous to an environment in the theory of open quantum systems \cite{Ana97}. This noise may lead to decoherence of correlation histories \cite{DCH}, i.e., to the classicalization of the effective description.

The QTP approach demonstrates that the different levels of correlation histories can be accessed by the measurement of $n$-detector joint probabilities.
Eq. (\ref{probdenr}) assigns to each initial state $|\psi\rangle$ of the field a hierarchy of joint  probability distributions  $P_n(z_1, z_2, \ldots, z_n)$.

 In classical probability theory, a hierarchy of correlation functions defines a classical stochastic process, if it satisfies the Kolmogorov additivity condition,
\bey
P_{n-1}(z_1, \ldots, z_{n-1}) = \int dz_n  P_n(z_1, z_2, \ldots, z_n) \label{Kolmadd}
\eey
Quantum probability distributions for sequential measurements do not satisfy this condition \cite{Ana06, Brunnt}. Hence, the violation of Eq. (\ref{Kolmadd}) is a genuine signature  of quantum dynamics; it cannot be reproduced by classical physics, including classical stochastic processes. It is rather different from the Leggett-Garg inequalities \cite{LeGa} that also refer to the behavior of quantum multi-time probabilities.
In contrast, if measurements on a quantum field approximately satisfy Eq. (\ref{Kolmadd}), then the measurement outcomes can be  simulated by a stochastic process with $n$-time probabilities given by the probability distributions (\ref{Kolmadd}). Then the generating functional $Z_{QTP}$ corresponds to a stochastic process, i.e., it is obtained as the functional Laplace transform of a classical stochastic probability measure.

Hence, the hierarchy $P_n(z_1, z_2, \ldots, z_n)$ provides a natural and unambiguous classicality criterion. Given the relation between QTP probabilities and QFT correlation functions, this criterion can be used to probe the information content of different levels for correlation histories. For example, the validity of Eq. (\ref{Kolmadd}) is necessary for deriving deterministic or classical stochastic dynamics for $P_1(z)$, i.e., for deriving the Boltzmann or the Boltzmann-Langevin equation. The failure of (\ref{Kolmadd}) means that the four-point correlation functions are too `quantum' to allow effective  classical  stochastic dynamics for the two-point correlation function.

Conversely, the failure of Eq. (\ref{Kolmadd}) can be used to provide a measure of irreducibly quantum information at the level $n = 2$. An example of such a measure is
\bey
S_Q = \int dz_2 \left|\int dz_1 P_2(z_1, z_2) - P_1(z_2)\right|.
\eey
A second informational quantity is the correlation of the probability distribution, i.e., a measure of the deviation  of $P_2(z_1, z_2)$ from $P_1(z_1) P_2(z_2)$. The joint probability density $P_2$ is then slaved to $P_1$. This information is typically quantified by  the correlation entropy
\bey
S_C = \int dz_1 dz_2 P_2(z_1, z_2) \ln \frac{P_2(z_1, z_2)}{P_1(z_1) P_1(z_2)}.
\eey
In general, the correlation entropy will contain information for both quantum correlations (if $S_Q \neq 0$) and classical stochastic ones. Indeed, $S_C$ may not have the usual properties of correlation entropy if $S_Q \neq 0$, and other measures that will distinguish will be more convenient. The third relevant informational quantity in Boltzmann's entropy (\ref{boltzent}), defined in terms of $P_1(z_1)$. These three quantities are the most important for describing the flow of information at the level of the 2PIEA.

 Hence, the QTP hierarchy functions as a registrar of information of the quantum system,   keeping track of how much information resides in what order, and how it flows from one order to another through the dynamics. There is a good potential for this scheme to systemize quantum information in QFT:    keeping track of the contents and the flow of information and   measuring the degree of coherence in a quantum system.   This conceptual scheme was suggested in Refs.\cite{Erice95, AnSav20} to explain black hole information loss, for instance.

\section{Conclusions}

We presented the QTP formalism for measurements in quantum fields and its  connections to the Closed-Time-Path description of QFT. These connections provide  a direct translation between the operational language of measurement theory (POVMs, effects, and so on) to the manifestly covariant description of QFT through functional methods.
 For example, we showed how one can express  POVMs for particle detection in terms of path-integrals. We believe that this is an important step towards the formulation of   a general theory of relativistic quantum information.

A key aspect of our approach is  the central role of the hierarchy of unequal-time correlation functions. POVMs for measurement are linear functional of such correlation functions. The very same correlation functions define the real causal dynamics  in the CTP approach, which are essential for the definition of thermodynamical observables and the construction of effective irreversible dynamics in non-equilibrium QFT.

Crucially, QFT correlation functions are covariant and causal objects. For this reason, we believe that a relativistic quantum information theory (QIT) that respects both causality and spacetime symmetry must define all informational quantities in terms of such correlation functions. This contrasts the standard approach of QIT  that is based on properties of single-time quantum states, like von Neumann entropy or entanglement.

A sound theoretical foundation for relativistic QIT is important for reasons that go beyond    theoretical coherence. Such a foundation is required for the description and design of
 quantum experiments in space \cite{Rideout, DSQL2} that will explore the effects of   non-inertial motion (acceleration, rotation) and gravity on quantum correlations, including entanglement. The predictions of a QFT measurement theory may be testable in experiments   that involve long separations or large relative velocities between detectors. \\

\noindent{\bf Acknowledgements}  This research was supported by a grant JSF-19-07-0001 from the Julian Schwinger Foundation.

\end{document}